\documentclass[12pt]{article}
\usepackage{graphicx,amssymb,bbold,bm}
\usepackage{amsmath,amsfonts,epsfig}
\usepackage{chessfss}
\usepackage{braket}
\usepackage[usenames]{color} %%%%%%%%%%%
\usepackage{wrapfig}
\usepackage{amsfonts,bm}
\usepackage{tikz}
\newcommand{\p}{\partial}
\newcommand{\pslash}{p\kern-1ex /}
\newcommand{\qslash}{q\kern-1ex /}
\newcommand{\lslash}{l\kern-1ex /}
\newcommand{\sslash}{s\kern-1ex /}
\newcommand{\kaslash}{k_a\kern-2ex /}
\newcommand{\kbslash}{k_b\kern-2ex /}
\newcommand{\Dslash}{{\cal D}\kern-1.5ex /}

\newcommand{\bc}{\overline{c}}

\newcommand{\beqa}{\begin{eqnarray}}
\newcommand{\eeqa}{\end{eqnarray}}

\newcommand{\bpm}{\begin{pmatrix}}
\newcommand{\epm}{\end{pmatrix}}
\newcommand{\bbm}{\begin{bmatrix}}
\newcommand{\ebm}{\end{bmatrix}}

\def\p{\partial}

\makeatletter

\@addtoreset{equation}{section}
\makeatother
\usepackage{subfig}
\begin{document}

%\voffset -0.7 true cm
%\hoffset 1.1 true cm
%\topmargin 0.0in
%\evensidemargin 0.0in
%\oddsidemargin 0.0in
%\textheight 8.6in
%\textwidth 7.1in
%\parskip 10 pt

\voffset -0.7 true cm
\hoffset 1.5 true cm
\topmargin 0.0in
\evensidemargin 0.0in
\oddsidemargin 0.0in
\textheight 8.6in
\textwidth 5.4in
\parskip 9 pt
 
\def\Tr{\hbox{Tr}}
\newcommand{\be}{\begin{equation}}
\newcommand{\ee}{\end{equation}}
\newcommand{\bea}{\begin{eqnarray}}
\newcommand{\eea}{\end{eqnarray}}
\newcommand{\beas}{\begin{eqnarray*}}
\newcommand{\eeas}{\end{eqnarray*}}
\newcommand{\nn}{\nonumber}
\font\cmsss=cmss8
\def\C{{\hbox{\cmsss C}}}
\font\cmss=cmss10
\def\bigC{{\hbox{\cmss C}}}
\def\scriptlap{{\kern1pt\vbox{\hrule height 0.8pt\hbox{\vrule width 0.8pt
  \hskip2pt\vbox{\vskip 4pt}\hskip 2pt\vrule width 0.4pt}\hrule height 0.4pt}
  \kern1pt}}
\def\ba{{\bar{a}}}
\def\bb{{\bar{b}}}
\def\bc{{\bar{c}}}
\def\bphi{{\Phi}}
\def\Bigggl{\mathopen\Biggg}
\def\Bigggr{\mathclose\Biggg}
\def\Biggg#1{{\hbox{$\left#1\vbox to 25pt{}\right.\n@space$}}}
\def\n@space{\nulldelimiterspace=0pt \m@th}
\def\m@th{\mathsurround = 0pt}

\begin{titlepage}
\begin{flushright}
{\small OU-HET-1028}
 \\
\end{flushright}

\begin{center}

\vspace{5mm}
{\Large \bf {Conformally invariant averaged null energy}} \\[3pt] 
\vspace{1mm}
{\Large \bf  { condition from AdS/CFT}} \\[3pt]  
%\vspace{1mm}
%{\Large \bf  {the bulk vs boundary causality}}  
%{\Large \bf  {from gauge/gravity duality}} \\[3pt] 
%
%{\Large \bf { {What boundary metric can}} \\[3pt] 
%\vspace{1mm}
%{\Large \bf  { have holographic bulk dual?}}}  

\vspace{6mm}
%\vspace{10mm}
%\vspace{8mm}

\renewcommand\thefootnote{\mbox{$\fnsymbol{footnote}$}}
Norihiro Iizuka${}^{\textsymking}$, 
Akihiro Ishibashi${}^{\textsymbishop}$ and 
Kengo Maeda${}^{\textsymqueen}$
%Norihiro Iizuka${}^a$, 
%Akihiro Ishibashi${}^b$ and 
%Kengo Maeda${}^c$

\vspace{3mm}

${}^{\textsymking}${\small \sl Department of Physics, Osaka University} \\ 
{\small \sl Toyonaka, Osaka 560-0043, JAPAN}

${}^{\textsymbishop}${\small \sl Department of Physics and Research Institute for Science and Technology,} \\   
{\small \sl Kindai University, Higashi-Osaka 577-8502, JAPAN} 
%\\ 

${}^{\textsymqueen}${\small \sl Faculty of Engineering, Shibaura Institute of Technology,} \\   
{\small \sl Saitama 330-8570, JAPAN} 
%\\ 

\vspace{4mm}

{\small \tt 
{iizuka at phys.sci.osaka-u.ac.jp}, {akihiro at phys.kindai.ac.jp},   \\
{maeda302 at sic.shibaura-it.ac.jp}
}

\end{center}

\vspace{3mm}

\noindent
We study the compatibility of the AdS/CFT duality with the bulk and boundary causality, and derive a {\it conformally invariant} averaged null energy condition (CANEC) for quantum field theories in $3$ and $5$-dimensional curved boundaries. This is the generalization of the averaged null energy condition (ANEC) in Minkowski spacetime to curved boundaries, where null energy is averaged along the null line with appropriate weight for conformal invariance.  For this purpose we take, as our guiding principle, the no-bulk-shortcut theorem of Gao and Wald, which essentially asserts that when going from one point to another on the boundary, one cannot take a ``shortcut through the bulk''. We also discuss the relationship between bulk vs boundary causality and the weak cosmic censorship.

\end{titlepage}

\setcounter{footnote}{0}
\renewcommand\thefootnote{\mbox{\arabic{footnote}}}

\tableofcontents
\newpage
%%%%%%%%%%%%%%%%%%%%%%%%%%%%%%%%%%%%%%%%%%%%%%%%%%%%%%%%%%%

%%%%%%%%%%%%%%%%%%%%%%%%
\section{Introduction} 
%%%%%%%%%%%%%%%%%%%%%%%%

Holography, or gauge/gravity correspondence \cite{Maldacena:1997re,Gubser:1998bc,Witten:1998qj} is probably the most mysterious yet profound duality we learned through string theory. Although two decades has passed from its discovery, many aspects remain to be understood better. 
One of the key questions which we would like to pursue is as follows; given the boundary theory, when and how the holographic bulk dual emerges. For studies along this line, see for examples \cite{Heemskerk:2009pn, ElShowk:2011ag}. 
Although all of the consistent boundary (conformal) quantum field theories could, in some sense, define their bulk dual, it is far from obvious whether the corresponding bulk theories would also support the Einstein gravity\footnote{Recently it is pointed out that causality constraints with large $N$ and a sparse spectrum might restrict the bulk theory to the Einstein gravity \cite{Camanho:2014apa, Afkhami-Jeddi:2016ntf}. It is interesting to investigate these more.}.  

To deepen our understanding of AdS/CFT correspondence along the direction mentioned above, 
in this paper we focus on the causal structures of both bulk and boundary, and their relationships. 
Assuming that the bulk dynamics is dominated by the Einstein gravity, we would like to 
ask that in order for the bulk/boundary correspondence to work, how both of bulk and boundary causal structures need to be consistent.  Suppose that the boundary geometry is given and corresponding bulk emerges. Let us consider any two points in the boundary. If there is a shortcut causal curve through the bulk that connects these two boundary points, i.e., 
the bulk curve that can travel faster than the one on the boundary, then such a boundary cannot support the holographic bulk geometry. This is because propagator through the bulk carries information faster than the boundary one, resulting in mismatch between the bulk and 
boundary correlators. In other words, this implies 
that the fastest causal curve which connects two boundary points on an achronal null curve is the boundary null geodesic. 
This ``no-bulk-shortcut'' property was proven by Gao-Wald \cite{Gao:2000ga} (see also \cite{Witten:2019qhl}) 
under the assumption that there is no causal pathology in the bulk, such as 
naked singularities or causality violating region. 

Given these observations, it appears reasonable to promote the Gao-Wald's no-bulk-shortcut property 
to one of the guiding principles. With these principles, we would like to understand which boundary metric 
can support the emergent bulk as Einstein gravity.
Then our main question, which is twofold, is the following: 

\begin{itemize}
\item[(i)] Given generic curved boundary metric, what kind of effects on the boundary theory does the no-bulk-shortcut property give rise to?

\item[(ii)] What kind of boundary field theory possibly admits a holographic bulk dual with bulk-shortcut?

\end{itemize} 

The key to answering these questions (i) and (ii) is the averaged null energy condition~(ANEC). 
In fact, concerning the first question (i),  in flat Minkowski boundary case, Kelly and Wall~\cite{Kelly:2014mra} 
derived an ANEC in the boundary theory from the no-bulk-shortcut condition.  
Note that in the flat Minkowski background case,  there is a completely field theoretic proof by~\cite{Faulkner:2016mzt, Hartman:2016lgu}. Therefore we expect that by considering curved boundary metrics, 
we may also be able to obtain a similar type of non-local null energy inequality along an achronal 
null geodesic segment in a class of strongly coupled field theories by using the AdS/CFT duality. 
As for the second question (ii), for such a boundary theory, 
we expect that if the bulk allows a short-cut, then it manifestly contradicts with the boundary causality and 
therefore that AdS/CFT duality fails to hold on such a boundary.   

In the Einstein gravity, ANEC is one of the most fascinating non-local null energy 
conditions because it is essential to prove the singularity theorem, topological censorship, 
and other important theorems in classical general relativity. 
In a conformally coupled free scalar field theory, however, ANEC can be violated in conformally flat 
spacetime~\cite{Urban:2009yt}~(see also Ref.~\cite{Visser:1994jb} for general curved spacetime). 
Recently, the violation of ANEC has also been shown in strongly coupled field theories in curved spacetime 
by using the AdS/CFT duality~\cite{Ishibashi:2019nby}.     
The key issue for both counter-examples of the ANEC is that local violation of the null energy condition at a point 
can be {\it enhanced via conformal transformations}. This is mainly due to the fact that ANEC is {\it not} conformally invariant. Therefore, ANEC can be violated in the conformally transformed spacetime.  
%\footnote{Flat Minkowski spacetime is conformally invariant background, therefore such negative null energy enhancement cannot occur there.}. 
Given these, in order to generalize ANEC to curved background boundary quantum field theories, it is natural to ask whether there is a 
{\it conformally invariant} averaged null energy condition, since such quantity can restrict the extent of the local  
negative null energy.

Motivated by the questions raised above, in this paper we shall study relationships between the bulk and boundary 
causal structures in the context of holography. 
Concerning the question (i), we consider $4$ (and $5$)-dimensional asymptotically AdS vacuum spacetime in which 
the conformal boundary is a static spatially compact universe, and derive a new {\it conformally invariant} averaged null energy condition (CANEC) along an achronal null geodesic segment in a class of strongly coupled field theories by using the AdS/CFT duality. As a special case, when the conformal boundary is a flat spacetime, our CANEC reduces to the usual ANEC, thus 
being consistent with the result of Kelly and Wall~\cite{Kelly:2014mra}.  
Our CANEC is written only in terms of quantities in the boundary such as Jacobi fields~\cite{HawkingEllis} 
associated with boundary null geodesics. 
We shall also consider $6$-dimensional asymptotically AdS vacuum bulk spacetime 
of which conformal boundary is a $5$-dimensional deformed static Einstein universe whose compact spatial sections are not symmetric but deformed sphere. Also in this case, we show that the no-bulk-shortcut property leads to CANEC in $5$-dimensional boundaries. 
Because of its conformal invariance, CANEC introduced in this paper would be more useful than ANEC to put restrictions 
on the behavior of boundary stress-energy tensors.  

As for the question (ii), we briefly discuss that if CANEC is violated on a given boundary theory, 
then the corresponding bulk must admit a bulk-shortcut, giving rise to inconsistency between the bulk and the boundary causality. 
%violate the weak cosmic censorship, forming a naked singularity. 
In this way, our boundary CANEC can be used as one of the criteria for having 
a consistent, healthy arena for AdS/CFT duality. 
% we consider $6$-dimensional asymptotically AdS vacuum bulk spacetime 
% of which conformal boundary is a $5$-dimensional deformed static Einstein universe whose compact spatial sections are not symmetric 
% but deformed sphere. When the deformation parameter is larger than a certain critical value, we find that there is 
% a bulk timelike curve near the AdS boundary which connects two points on an achronal null %curve on the boundary. 
%This implies that there must be a naked singularity in the bulk, according to the results of Ref.~\cite{Ishibashi:2019nby}. 
  
In the next section we explain basic formulas we use in the subsequent sections. 
In Sec.~\ref{sec:3}, using no-bulk-shortcut property, we derive, holographically, CANEC 
%conformally invariant averaged null energy condition (CANEC) 
for $3$-dimensional boundary, which is static Einstein universe with constant curvature. 
Then in Sec.~\ref{sec:4}, we generalize the previous results to the deformed compact boundary. 
We also show the conformal invariance of CANEC. 
In Sec.~\ref{sec:5}, we consider $5$-dimensional deformed compact boundary and %show that if deformation is large enough, we will obtain a bulk-short cut. 
derive CANEC holographically. 
Sec.~\ref{sec:summary} summarizes our results. 
In Appendix, we briefly discuss $4$-dimensional deformed 
boundary universe, which admits conformal anomalies.

%%%%%%%%%%%%%%%%%%%%%%
\section{Preliminaries}
\label{sec:2}
%%%%%%%%%%%%%%%%%%%%%%
We consider $d+1$-dimensional asymptotically AdS$_{d+1}$ spacetime $(M,\,g_{ab})$ with $d$-dimensional conformal timelike boundary $\p M$. 
We first provide our geometric conditions, including more detailed statement of the no-bulk-shortcut property, 
and then recall the formulas for a renormalized boundary stress energy tensor.

\subsection{No bulk-shortcut property} 
Let us consider an arbitrary pair of two points $p$ and $q$ 
on the conformal boundary ${\p M}$ 
connected by a null geodesic $\gamma$ that is lying entirely in the boundary ${\p M}$ and is 
achronal with respect to the boundary metric. 
If there is a timelike curve through the bulk $M$ that connects the boundary two points $p$ and $q$, 
then the entire spacetime $M\cup \p M$ is said to admit a {\em bulk-shortcut}.

When there is no such a bulk-shortcut, the null geodesic curve segment $\gamma$ from $p$ to $q$ is called 
the {\em fastest causal curve from $p$ to $q$} in the entire geometry. 
Within the Einstein gravity, it has been shown by Gao and Wald in Ref.~\cite{Gao:2000ga} that this is indeed the case 
for spacetimes satisfying some physically reasonable conditions, including the null energy condition, or more generally ANEC for the matter fields in the bulk. It was shown recently that the lack of this property can be connected with the violation of the weak cosmic censorship~\cite{Ishibashi:2019nby}. 
In the rest of this paper, when a spacetime under consideration does not admit any bulk-shortcut, we refer to such a spacetime as possessing the {\em no-bulk-shortcut property}.

If an asymptotically AdS spacetime does not possess the no-bulk-shortcut property, then while a boundary field theory correlator must 
have its support on and inside the boundary lightcone (e.g., the one emanating from a boundary point $p$ with the other boundary point $q$ on it), the corresponding bulk correlator 
would involve some correlation between $p$ and some other boundary point which is strictly to the past of $q$ with respect to the boundary metric. 
Thus it gives rise to a causality violation from the boundary field theory viewpoint. 
Therefore, in that case, one would not be able to construct a consistent holographic model. 
In view of this, it is reasonable to regard the no-bulk-shortcut property as one of the guiding principles for a sensible choice of geometry $(M,g_{ab})$ 
for which the AdS/CFT duality works properly.

In addition, we also impose the following conditions,  
\begin{enumerate}
\item The bulk spacetime satisfies the Einstein equations 
\begin{align}
R_{ab}-\frac{1}{2}Rg_{ab}=-\Lambda g_{ab}+\kappa^2 T_{ab}, \qquad \Lambda=\frac{d(d-1)}{2l^2}, 
\end{align}
where $l$ is the length of the negative cosmological constant, $\kappa^2$ is the gravitational 
constant, and $T_{ab}$ is the stress-energy tensor in the bulk.   
\item Near the boundary $\p M$, $T_{ab}=0$. 
\end{enumerate}
Physically the second condition implies that all the matter fields do not extend to spatial infinity 
and there is no matter radiation reaching the null infinity, $\p M$. The second condition may be replaced 
with the condition that bulk matter fields decay sufficiently quickly toward the boundary. 

\subsection{Boundary stress-energy tensor} 
For later convenience, we adopt the following   
$d+1$-dimensional coordinates in the bulk, 
\begin{align}
\label{FG_coordinates}
& ds^2=\frac{\hat{g}_{ab}dx^a dx^b}{z^2}=\frac{dz^2+\hat{g}_{\mu\nu}(z,\,x)dx^\mu dx^\nu}{z^2}, 
\end{align}  
where $z=0$ corresponds to the conformal boundary with $\hat{g}_{\mu\nu}(z=0,\,x)$ being the boundary metric. 
Since any conformal transformation does not change the causal structure, 
we consider the conformally transformed metric $\hat{g}_{ab}dx^a dx^b$ below so that the boundary metric becomes regular.    

In general, the metric $\hat{g}_{\mu\nu}$ can be expanded as 
\begin{align}
\label{Fefferman-Graham_expansion}
\hat{g}_{\mu\nu}(z,\,x)=\sum_{n=0}^\infty g_{(n)\mu\nu}(x)z^n+ z^d\ln z^2\, h_{\mu \nu}(x), 
\end{align} 
where the logarithmic term appears only for $d$ = even, {\it i.e.,} even boundary dimensions, and $g_{(2k+1)\mu\nu}=0$ for any integer $k$ 
satisfying $0\le 2k+1<d$~\cite{deHaro:2000vlm}. 
The coordinate system $(z,x^\mu)$ in which the metric takes the form of (\ref{FG_coordinates}) with the expansion (\ref{Fefferman-Graham_expansion}) 
is called the Fefferman-Graham coordinate system. 
For a given $g_{(0)\mu\nu}$ metric, the subleading terms  
$g_{(2)\mu\nu}$ and $g_{(4)\mu\nu}$~($d>4$) are given as~\cite{deHaro:2000vlm}\footnote{We follow the Wald's textbook~\cite{Wald} for the notation of the curvature. 
Therefore, the sign of $g_{(2)\mu\nu}$ and of partial $g_{(4)\mu\nu}$~($d>4$) are different from the one~\cite{deHaro:2000vlm}.} 
\begin{align}
\label{coeff_2}
g_{(2)\mu\nu}= - \frac{1}{d-2}\left(\hat{R}_{\mu\nu}-\frac{1}{2(d-1)}\hat{R}\,g_{(0)\mu\nu} \right), 
\end{align}
\begin{align}
\label{coeff_4}
& g_{(4)\mu\nu}=\frac{1}{d-4}\Biggl(\frac{1}{8(d-1)}\nabla_\mu\nabla_\nu \hat{R}-\frac{1}{4(d-2)}\nabla^\alpha\nabla_\alpha 
\hat{R}_{\mu\nu}
\nonumber \\
&+\frac{1}{8(d-1)(d-2)}g_{(0)\mu\nu}\nabla^\alpha\nabla_\alpha \hat{R}
-\frac{1}{2(d-2)}\hat{R}^{\alpha\beta}\hat{R}_{\mu\alpha\nu\beta}+\frac{d-4}{2(d-2)^2}{\hat{R}_\mu}^{\,\,\,\alpha} 
\hat{R}_{\alpha\nu} \nonumber \\
&+\frac{1}{(d-1)(d-2)^2}\hat{R}\hat{R}_{\mu\nu}
+\frac{1}{4(d-2)^2}\hat{R}^{\alpha\beta}\hat{R}_{\alpha\beta}g_{(0)\mu\nu}-\frac{3d}{16(d-1)^2(d-2)^2}\hat{R}^2g_{(0)\mu\nu}\Biggr), 
\end{align}
where $\hat{R}_{\mu\nu}$ and $\hat{R}$ are the Ricci tensor and the Ricci scalar of the conformal boundary metric $g_{(0)\mu\nu}$, 
respectively.  

All the subleading coefficients $g_{(n)\mu\nu}$ with $n<d$ are determined from the boundary 
metric $g_{(0)\mu\nu}$ and the coefficient $g_{(d)\mu\nu}$ corresponds to the renormalized 
stress energy tensor on the boundary,   
\begin{align}
\label{renormalized_stress_energy}
\braket{{T}_{\mu\nu}}=\frac{dl^{d-1}}{16\pi G_{d+1}}g_{(d)\mu\nu}+X_{\mu\nu}, 
\end{align}
where $X_{\mu\nu}$ represents the gravitational conformal anomaly, which vanishes for the odd-dimensional boundaries. 
In the following sections, for simplicity, we shall restrict our attention to such odd-dimensional boundaries.   
We briefly discuss the even-dimensional ($d=4$) boundary case in Appendix A.

%%%%%%%%%%%%%%%%%%%%%%%%%%%%%%
\section{$d=3$ boundary spacetimes with constant spatial curvature}
\label{sec:3}
%%%%%%%%%%%%%%%%%%%%%%%%%%%%%%
In this section, we start by considering $d=3$ static boundary spacetimes with a constant spatial curvature. 
Such a boundary spacetime is naturally derived from typical asymptotically AdS 
spacetimes. For example, the boundary spacetime of the asymptotically global AdS spacetime is 
the static Einstein universe with positive spatial curvature, for which we derive a null energy inequality 
from the no-bulk-shortcut property. For the case with negatively curved spatial section, 
on the other hand, the no-bulk-shortcut condition appears to be always satisfied, as far as analyzing a certain bulk region near the boundary 
where the Fefferman-Graham expansion works.

%%%%%%%%%%%%%%%%%%%%%%%%%%%%%%%%%%%%%
\subsection{Static compact universe}
%%%%%%%%%%%%%%%%%%%%%%%%%%%%%%%%%%%%%
First, we start with $3$-dimensional Einstein static universe with constant positive spatial 
curvature, as our conformal boundary metric ${g}_{(0) \mu\nu}$. The metric is written by 
\begin{align}
\label{3dboundarymetric}
& ds_3^2=-dt^2+d\rho^2+\sin^2 \rho \,d\varphi^2=-2dUdV+\sin^2 \left(\frac{V-U}{\sqrt{2}} \right) d\varphi^2, \nonumber \\
& U=\frac{t-\rho}{\sqrt{2}}, \qquad V=\frac{t+\rho}{\sqrt{2}}, 
\end{align} 
where $0\le \rho\le \pi$, $\varphi$ the angular coordinate with $0\le \varphi\le 2\pi$, and $U$, $V$ are null 
coordinates. Consider the boundary null geodesic segment $\gamma$ along $U=0$, $\varphi=0$ with the tangent vector $l^\mu=(\p_V)^\mu$ 
from a boundary point $p:(z, U, V)=(0,0,0)$ at the, say, south pole to a point $q:(0, 0, V_0)=(0,0, \sqrt{2}\pi)$ at the corresponding north 
pole. This null geodesic segment is achronal on the boundary, and the null-null component of the Ricci tensor becomes 
\begin{align}
\label{Rvv}
\hat{R}_{\mu\nu} l^\mu l^\nu  =\frac{1}{2}.  
\end{align}

%%%%%%%%%%%%%%%%
\begin{figure}[tbp]
\begin{center}
\includegraphics[width=.5\textwidth]{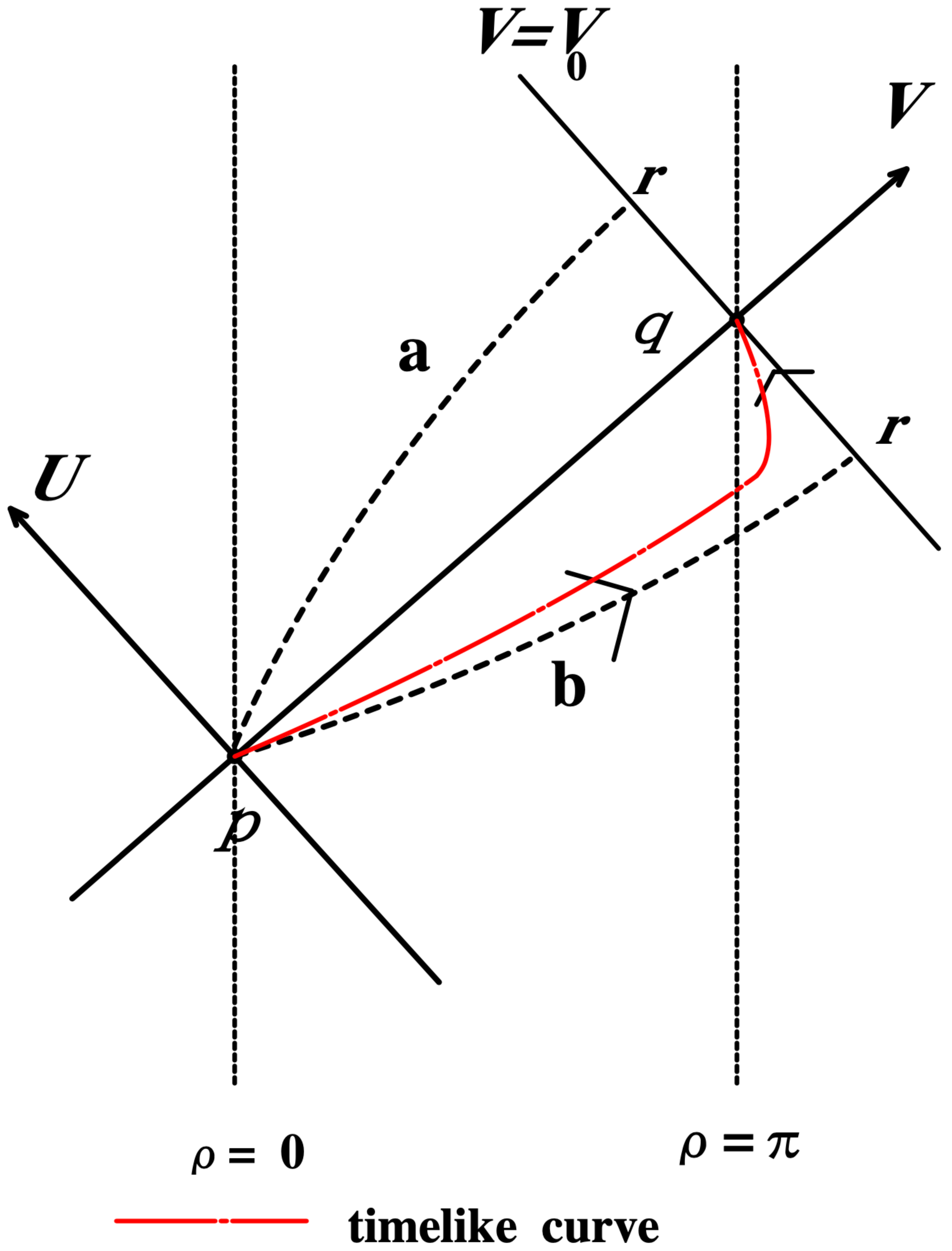}
\vspace{2mm}
\caption{\label{double_null} The dashed curves represent two possible cases of the bulk causal curve $\xi$ 
that connects two boundary points, the one $p$ at $U=V=0$ and the other $r$ on $V=V_0$. 
(i) When $r$ is in the future of $q$ along $V=V_0$, the corresponding bulk curve $\xi$ is a delayed curve with respect to the boundary null 
geodesic $\gamma$ with $U>0$ at $r$ as depicted by the dashed curve {\bf a}.  
(ii) When $r$ is in the past of $q$ along $V=V_0$, then $\xi$ becomes a superluminal curve with respect to $\gamma$ with $U<0$ at $r$ 
as depicted by the dashed curve {\bf b}.  
In the latter case (ii), the broken null curve from $p$ to $q$ through $r$ implies that there is a timelike curve~(red broken curve) from $p$ to $q$.} 
\end{center}
\end{figure}
%%%%%%%%%%%% 

Let us consider the bulk causal curves near the boundary null geodesic $\gamma$, which connect two boundary 
points $p$ and $r$ with $r$ being on the null line $V=V_0$ (which is a null geodesic with the tangent $\p_U$) 
through the point $q$~(see, Fig.~\ref{double_null}). We denote one of such bulk causal curves by $\xi$. 
Now if $r$ is in the future of $q$, then the coordinate value of $U$ at $r$ is positive, and $\xi$ is just like 
the path {\bf a} in Fig.~\ref{double_null}. 
On the other hand, if $r$ is to the past of $q$, then $U$ at $r$ is negative and $\xi$ is like the path {\bf b} in Fig.~\ref{double_null}, 
being a part of the broken null curve from $p$ to $q$ through $r$~\footnote{Broken causal curve contains a point in which the tangent vector 
is discontinuous.}. In this case, $p$ and $q$ are also joined by a timelike curve, according to Proposition 4.~5.~9 in 
Ref.~\cite{HawkingEllis}, and therefore the no-bulk-shortcut condition is not satisfied.

In order to see in more detail what should happen under the no-bulk-shortcut condition, let us examine the causal curve $\xi$ 
which gives the minimum of $U$ at $r$. The tangent vector $K$ is generally expanded as 
\begin{align}
\label{bulk_causal_curve}
K^a & =\left(\frac{dz}{d\lambda},\, \frac{dU}{d\lambda},\,\frac{dV}{d\lambda},\,\frac{d\varphi}{d\lambda} \right)=
\left(\frac{dz(\lambda)}{d\lambda}, \, K^U,\, 1,\, 0\right), \nonumber \\
z &  =\epsilon z_1+\epsilon^2 z_2+\cdots, \\
K^U & =\epsilon^2 \frac{du_2}{d\lambda}+\epsilon^3 \frac{du_3}{d\lambda}+\cdots,  \nonumber 
\end{align}
where $\epsilon$ is an arbitrary small parameter and $\lambda~(0\le \lambda\le \sqrt{2}\pi)$ is the parameter of 
the causal curve. Since $p$ and $r$ are boundary points, $z$ must satisfy  
\begin{align}
\label{boundary_cond_z}
z(0)=z(\sqrt{2}\pi)=0. 
\end{align}
By the fact that $\xi$ is the non-spacelike curve, $\hat{g}_{ab}K^aK^b\le 0$. Using Eq. \eqref{coeff_2} and  \eqref{3dboundarymetric} - \eqref{bulk_causal_curve}, 
 we obtain  up to $O(\epsilon^2)$ as 
\begin{align}
\label{u2_IE_positive_con}
\frac{du_2}{d\lambda}\ge \frac{1}{2}\left(\dot{z}_1^2-\hat{R}_{\mu\nu} l^\mu l^\nu z_1^2   \right)
=\frac{1}{2}\left(\dot{z}_1^2-\frac{1}{2}z_1^2   \right)\ge 0, 
\end{align}
where a dot means the derivative with respect to $\lambda$ and the equality holds only if $K$ is a null vector. 

By integrating Eq.~(\ref{u2_IE_positive_con}), 
$U(\sqrt{2}\pi)$ must satisfy 
\begin{align}
\label{Delta_U_positive_con}
 U(\sqrt{2}\pi)\ge {\cal I} & := \frac{\epsilon^2}{2}\int^{\sqrt{2}\pi}_0 \left(\dot{z}_1^2-\frac{z_1^2}{2}\right)d\lambda+O(\epsilon^3),   \\
\qquad z_1(0) &=z_1(\sqrt{2}\pi)=0,  
\end{align}  
where the second equation is the boundary condition~(\ref{boundary_cond_z}).   
So, if there were $z_1(\lambda)$ with a negative value ${\cal I}$, then there were a bulk null curve with a negative 
$U(\sqrt{2}\pi)$~(note that the equality holds only if $K$ is null). This would contradict the no-bulk-shortcut condition.   

Applying the variational principle with respect to $z_1$ to ${\cal I}$, one finds the equation of motion for $z_1$ as 
\begin{align}
\label{Eq_min_positive_con}
\ddot{z}_1+\frac{1}{2}z_1=0, 
\end{align}
which gives the minimum\footnote{To see minimum, note that if $z_1$ oscillates rapidly, then clearly that increases $ {\cal I}$.} of ${\cal I}$, and the solution is 
\begin{align}
\label{sol:z_1_positive_con}
z_1=\sin\frac{\lambda}{\sqrt{2}}, 
\end{align} 
where the amplitude is set equal to unit without loss of generality. Substituting this into the r.~h.~s. of Eq.~(\ref{Delta_U_positive_con}), 
we find that ${\cal I}=0$, implying that the bulk null causal curve obeying Eq.~(\ref{sol:z_1_positive_con}) ends on the boundary 
point $q$ at the same moment with the null geodesic $\gamma$, up to $O(\epsilon^2)$. 
  
At third order in $\epsilon$, imposing that $\xi$ is the non-spacelike curve, one obtains   
\begin{align}
\label{u3_IE_positive_con}
\frac{du_3}{d\lambda}\ge \left(\dot{z}_1\dot{z}_2-\frac{1}{2}z_1z_2+ \frac{1}{2} z_1^3\,g_{(3)\mu\nu}\,l^\mu l^\nu   \right), \qquad 
z_2(0)=z_2(\sqrt{2}\pi)=0. 
\end{align}  
Integration of the inequality from $\lambda=0$ to $\lambda=\sqrt{2}\pi$ yields 
\begin{align}
\label{U3_positive_con}
U(\sqrt{2}\pi)\ge \frac{\epsilon^3}{2} \int^{\sqrt{2}\pi}_{0}z_1^3\,g_{(3)\mu\nu}\,l^\mu l^\nu d\lambda  
=\frac{\epsilon^3}{2} \int^{\sqrt{2}\pi}_{0}\left(\sin\frac{\lambda}{\sqrt{2}}\right)^3\,g_{(3)\mu\nu}\,l^\mu l^\nu d\lambda
\ge 0, 
\end{align}
where we used the integration by part under the boundary condition $z_2(0)=z_2(\sqrt{2}\pi)=0$ and Eq.~(\ref{Eq_min_positive_con}). 
For the odd-dimensional case, $g_{(3)\mu\nu}$ corresponds to the renormalized stress-energy tensor on the boundary by 
Eq.~(\ref{renormalized_stress_energy}). Therefore the inequality~(\ref{U3_positive_con}) reduces to the null energy 
inequality 
\begin{align}
\label{NEC_positive_con}
\int^{\sqrt{2}\pi}_{0}\left(\sin\frac{\lambda}{\sqrt{2}}\right)^3\, \braket{T_{\mu\nu}}l^\mu l^\nu d\lambda\ge 0, 
\end{align}
which restricts the extent of the negative value of the null energy $ \braket{T_{\mu\nu}}l^\mu l^\nu$. 
As shown later, the weight function $z_1^3 = \left(\sin \frac{\lambda}{\sqrt{2}} \right)^3$ corresponds to the cube of the Jacobi field $\eta$ of the null geodesic congruence of the null geodesic $\gamma$. We will show in next section that in order that \eqref{NEC_positive_con} is conformally invariant,  it is crucial to include this weight function.    

%%%%%%%%%%%%%%%%%%%%%%%%%%% 
\subsection{$d=3$ hyperbolic universe}    
%%%%%%%%%%%%%%%%%%%%%%%%%%%%  
Next, we consider $d=3$ static boundary spacetime with a negative spatial constant curvature. Our conformal boundary metric ${g}_{(0) \mu\nu}$  is given by 
\begin{align}
\label{negative_3_metric}
& ds_3^2=-dt^2+d\rho^2+\sinh^2 \rho \,d\varphi^2=-2dUdV+\sinh^2  \left(\frac{V-U}{\sqrt{2}} \right) \,d\varphi^2,  \nonumber \\
& \hat{R}_{\mu\nu} l^\mu l^\nu=-\frac{1}{2}. 
\end{align} 
As in the positive constant curvature case, we consider the bulk causal curve $\xi$ with tangent vector~(\ref{bulk_causal_curve}), 
and we obtain 
\begin{align}
\label{Delta_U_negative_con}
U(\sqrt{2}\pi)\ge {\cal I} &:=\frac{\epsilon^2}{2}\int^{\sqrt{2}\pi}_0 \left(\dot{z}_1^2+\frac{z_1^2}{2}\right)d\lambda+O(\epsilon^3), \\
\qquad z_1(0) &=z_1(\sqrt{2}\pi)=0  
\end{align}  
at $O(\epsilon^2)$. 
Hence $U$ at point $r$ is strictly positive unless $z_1\equiv 0$ at O($\epsilon^2$). 
Since $z_1 = 0$, one can see that contribution at the next order O($\epsilon^3$), which is analogous to Eq.~\eqref{u3_IE_positive_con} vanishes, independently of the subleading behavior of $g_{(3)\mu\nu}l^\mu l^\nu$. 
This is consistent with the no-bulk-shortcut property, and implies that one cannot derive 
the null energy inequality which restricts the extent of the negative value of the null energy $ \braket{T_{\mu\nu}}l^\mu l^\nu$.  
Actually, one can see this fact from $4$-dimensional hyperbolic AdS black hole solution~\cite{Emparan:1999gf}, which is dual to the 
thermal state of the boundary theory in the background spacetime~(\ref{negative_3_metric}). 
In this case, the mass can be negative, and accordingly $g_{(3)\mu\nu}l^\mu l^\nu$ can become a negative constant. This fact leads us to 
speculate that the averaged null energy condition  
can be derived from the no-bulk-shortcut property for spatially compact universe 
with positive (or zero) spatial curvature only\footnote{Note that this does not prohibit boundary spacetime to have negative curvature locally.}.

%%%%%%%%%%%%%%%%%%%%%%%%%%%%%%%%%%%%%%%%%%
\section{Conformally invariant averaged null energy condition in $d=3$ deformed compact universe}
\label{sec:4} 
%%%%%%%%%%%%%%%%%%%%%%%%%%%%%%%%%%%%%%%%%%
In this section we extend the results in the previous section to a more general class of $d=3$ dimensional boundary spacetimes. 
In particular, we consider $3$-dimensional spatially compact universe with non-constant spatial curvature, and write down conformally invariant averaged null energy condition (CANEC) in terms of the Jacobi field of the null geodesic congruence of the null geodesic $\gamma$. 
We assume that the conformal boundary metric is static and circularly symmetric. 
%%%%%%%%%%%%%%%%%%%%%%%%%%%%%%%%%%%%%%%%
\subsection{Conformally invariant averaged null energy condition}
%%%%%%%%%%%%%%%%%%%%%%%%%%%%%%%%%%%%%%%%

We consider the following conformal boundary metric ${g}_{(0) \mu\nu}$, 
\begin{align}
ds_3^2=g_{(0)\mu\nu}dx^\mu dx^\nu=e^{\psi}(-dt^2+d\rho^2)+\mu \, d\phi^2=-2e^{\psi}dUdV+\mu \, d\phi^2,  
\end{align}  
where $\psi$ and $\mu~(\mu\ge 0)$ are functions of $\rho~(0\le \rho\le \pi)$. 
As in the previous section, $\mu$ vanishes 
at both the south pole $\rho=0$ and the north pole $\rho=\pi$. 
This is more general than the metric \eqref{3dboundarymetric}. 
Let us consider a future-directed null geodesic $\gamma$ 
along $U=0$ and $\varphi=0$ from the south pole point, $p$, to the north pole point, $q$ along the boundary.   
As easily checked, 
\begin{align}
\label{null_geodesic}
l=e^{-\psi}  \p_V  
\end{align}
is the tangent vector of the boundary null geodesic, which is achronal between $p$ and $q$, satisfying $l^\nu \nabla_\nu l^\mu = 0$. 
Let us denote the values of the affine parameter $\lambda$ at $p$, $q$, by $\lambda_-$, $\lambda_+$, respectively, 
and also denote the value of coordinate $V$ at $q$ by $V_0$. 
The Raychaudhuri equation of the null geodesic congruences with tangent vector $l$ along the 3-dimensional boundary is  
\begin{align}
\label{Raycha_general_3}
\frac{d\theta}{d\lambda}=-\theta^2-\hat{R}_{\mu\nu}l^\mu l^\nu,  
\end{align} 
where $\theta$ is the expansion of the null geodesics. By defining 
$\theta:=(d\eta/d\lambda)/\eta$, Eq.~(\ref{Raycha_general_3}) is reduced to 
\begin{align}
\label{Jacobi_general_3}
\frac{d^2\eta}{d\lambda^2}=-\hat{R}_{\mu\nu}l^\mu l^\nu\,\eta.  
\end{align}
The function $\eta$, which is called the Jacobi field~\cite{HawkingEllis}, represents separation of points {\it between the two 
adjacent null geodesics on the boundary}. This vanishes at both the south and north poles, 
\begin{align}
\eta(\lambda_-)=\eta(\lambda_+)=0, 
\end{align} 
which are conjugate to each other, {\it i.e.,} they are focal points. 
Note that this Jacobi field is defined on the boundary only. 

Following the procedure in the previous section, we consider the bulk causal curves $\xi$ near the null geodesic $\gamma$ 
which connects two boundary points $p$ and $r$. Here, $r$ is a point on the $V=V_0$ null geodesic line along 
$\p_U$ through $q$. The tangent vector $K$ is written 
\begin{align}
K^a=\left(\frac{dz}{d\lambda},\, \frac{dU}{d\lambda},\,\frac{dV}{d\lambda},\,\frac{d\phi}{d\lambda} \right)
=(\dot{z}, K^U, K^V, 0)=(\dot{z}, K^U, e^{-\psi}, 0),   
\end{align}
where $z$ and $K^U$ can be expanded as Eq.~(\ref{bulk_causal_curve}) as a series in the small parameter $\epsilon$.  

At $O(\epsilon^2)$, the inequality $\hat{g}_{ab}K^a K^b\le 0$ reduces to 
\begin{align}
\frac{d{u_2}}{d\lambda}\ge \frac{1}{2} \left( \dot{z}_1^2-z_1^2\hat{R}_{\mu\nu}l^\mu l^\nu \right) . 
\end{align}
Integrating the above inequality from $\lambda=\lambda_-$ to $\lambda=\lambda_+$, we obtain 
\begin{align}
\label{delta_r}
U(\lambda_+)=\epsilon^2\int^{\lambda_+}_{\lambda_-}\frac{d{u_2}}{d\lambda}d\lambda
\ge {\cal I}:= \frac{\epsilon^2}{2} \int^{\lambda_+}_{\lambda_-}\left(\dot{z}_1^2-\hat{R}_{\mu\nu}l^\mu l^\nu z_1^2 \right)d\lambda,  
\end{align}
up to $O(\epsilon^2)$. Under the boundary conditions 
\begin{align}
\label{bc_general}
z(\lambda_\pm)=0, 
\end{align}
the r.~h.~s. of Eq.~(\ref{delta_r}) is minimized when 
\begin{align}
\label{general_z1}
\ddot{z}_1=-\hat{R}_{\mu\nu}l^\mu l^\nu z_1. 
\end{align} 
Therefore, $U(\lambda_+) \ge 0$ up to $O(\epsilon^2)$, where equality holds only for the null curve.

At $O(\epsilon^3)$, $\hat{g}_{ab}K^a K^b\le 0$ yields  
\begin{align}
\frac{d{u_3}}{d\lambda}\ge \dot{z}_1\dot{z}_2-z_1z_2\hat{R}_{\mu\nu}l^\mu l^\nu + \frac{1}{2} z_1^3\,g^{(3)}_{\mu\nu}\,l^\mu l^\nu. 
\end{align}
Integrating by part from $\lambda_-$ to $\lambda_+$ under the condition~(\ref{bc_general}), 
one obtains 
\begin{align}
& U(\lambda_+)\ge \frac{\epsilon^3}{2} \int^{\lambda_+}_{\lambda_-}z_1^3g^{(3)}_{\mu\nu}l^\mu l^\nu d\lambda,   
\end{align} 
where the equality holds for the bulk null curve. 
From the no-bulk-shortcut property, $U(\lambda_+)$ should be non-negative, and hence 
we derive the following inequality 
\begin{align}
\int^{\lambda_+}_{\lambda_-}z_1^3g^{(3)}_{\mu\nu}l^\mu l^\nu d\lambda \ge 0 \quad \Longleftrightarrow \quad 
\int^{\lambda_+}_{\lambda_-}z_1^3 \braket{T_{\mu\nu}}l^\mu l^\nu d\lambda\ge 0 \,, 
\end{align}
for the renormalized stress-energy tensor~(\ref{renormalized_stress_energy}) with vanishing $X_{\mu\nu}$. 
Note that Eq.~(\ref{general_z1}) is equivalent to Eq.~(\ref{Jacobi_general_3}) in the boundary theory. Therefore the above 
inequality is described in terms of the Jacobi field $\eta$ as  
\begin{align}
\label{Null_inequality_general}
\int^{\lambda_+}_{\lambda_-}\eta^3  \braket{T_{\mu\nu}}   l^\mu l^\nu d\lambda\ge 0. 
\end{align}
It is worth noting that the final inequality \eqref{Null_inequality_general}  is written in terms of only boundary quantities,  
independent of the bulk quantities.

%%%%%%%%%%%%%%%%%%%%%%%%%%%%%%%%%%%%%%
\subsection{Conformal invariance}
%%%%%%%%%%%%%%%%%%%%%%%%%%%%%%%%%%%%%%
We show that the inequality (\ref{Null_inequality_general}) is invariant under the conformal transformation of the boundary metric. 
Under the conformal transformation, 
\begin{align}
\label{conformal}
\tilde{g}_{(0)\mu\nu}=\Omega^2g_{(0)\mu\nu}. 
\end{align} 
the Ricci tensor, the null vector, the affine parameter, and the Jacobi field are transformed as 
\begin{align}
\label{conformal_t}
& \hat{R}_{\mu\nu}=\tilde{R}_{\mu\nu}+\Omega^{-1}\tilde{\nabla}_\mu \tilde{\nabla}_\nu\Omega
+2\Omega^{-2}\tilde{\nabla}^\alpha\Omega\tilde{\nabla}_\alpha\Omega\, \tilde{g}_{\mu\nu}
+\Omega^{-1}\tilde{\nabla}^\alpha\tilde{\nabla}_\alpha\Omega\, \tilde{g}_{\mu\nu}, \nonumber \\
& \tilde{l}^\mu=\frac{1}{\Omega^2}l^\mu ,   \qquad \frac{d\tilde{\lambda}}{d\lambda}=\Omega^2 , \qquad 
\tilde{\eta}=\Omega\, \eta  
\end{align}
where $\tilde{\nabla}_\mu$ is the covariant derivative with respect to $\tilde{g}_{\mu\nu}$ and $\tilde{l}$ is 
the tangent vector of the null geodesic on the conformal metric with the affine parameter $\tilde{\lambda}$. 
The transformation of the affine parameter can be understood from $l^\mu \nabla_\mu = \frac{d}{d \lambda}$, and the one of the Jacobi field can be understood since it represents the separation distance of adjacent null geodesics, which will be multiplied by $\Omega$, under the metric rescaling \eqref{conformal}.  

The l.~h.~s. of Eq.~(\ref{Jacobi_general_3}) is rewritten in terms of the quantities on the conformal metric as
\begin{align}
\frac{d^2\eta}{d\lambda^2}=\frac{d}{d\lambda}\left(\Omega^2\frac{d\eta}{d\tilde{\lambda}}\right)
=-\Omega^2\frac{d^2\Omega}{d\tilde{\lambda}^2}\tilde{\eta}+\Omega^3\frac{d^2\tilde{\eta}}{d\tilde{\lambda}^2} . 
\end{align}
On the other hand, the r.~h.~s. of Eq.~(\ref{Jacobi_general_3}) is 
rewritten in terms of the quantities on the conformal metric as
\begin{align}
 \hat{R}_{\mu\nu}l^\mu l^\nu\, \eta &=(\Omega^3 \tilde{R}_{\mu\nu}\tilde{l}^\mu\tilde{l}^\nu
+\Omega^2\tilde{l}^\mu\tilde{l}^\nu\tilde{\nabla}_\mu \tilde{\nabla}_\nu\Omega)\tilde{\eta} \nonumber \\
&=\left[\Omega^3 \tilde{R}_{\mu\nu}\tilde{l}^\mu\tilde{l}^\nu
+\Omega^2\tilde{l}^\mu\tilde{\nabla}_\mu(\tilde{l}^\nu \tilde{\nabla}_\nu\Omega)\right]\tilde{\eta} \nonumber \\
&=\left[\Omega^3 \tilde{R}_{\mu\nu}\tilde{l}^\mu\tilde{l}^\nu+\Omega^2\frac{d^2\Omega}{d\tilde{\lambda}^2}\right]\tilde{\eta},   
\end{align}
where we used the fact that $\tilde{l}^\mu$ is null vector and $\tilde{l}^\mu\tilde{\nabla}_\mu \tilde{l}^\nu=0$.  
Therefore, starting from 
\eqref{Jacobi_general_3}, after conformal transformation, we obtain 
\begin{align}
\frac{d^2\tilde{\eta}}{d\tilde{\lambda}^2}=-\tilde{R}_{\mu\nu}\tilde{l}^\mu\tilde{l}^\nu \tilde{\eta}. 
\end{align} 

Note that the renormalized stress-energy tensor~(\ref{renormalized_stress_energy}) transforms under the 
conformal transformation as\footnote{Since our boundary is odd-dimensional, there is no conformal anomaly.} 
\begin{align}
\braket{\tilde{T}_{\mu\nu}}=\Omega^{-1} \braket{T_{\mu\nu}}
\end{align}
for $d=3$~\cite{deHaro:2000vlm}. Combining this fact with the transformation Eq.~(\ref{conformal_t}), one finds the inequality (\ref{Null_inequality_general}) is conformally invariant. 

The achronal averaged null energy condition~\cite{Graham:2007va} in curved spacetime is clearly not conformally invariant. 
Therefore, as shown in~\cite{Ishibashi:2019nby}, for choosing a suitable conformal factor, it can be always violated, provided that 
the null energy condition is locally violated on the boundary theory. In the framework of the conformal boundary theory, 
all physical quantities should be described in a conformally invariant way. In this sense, Eq.~(\ref{Null_inequality_general}), 
the conformally invariant averaged null energy condition (CANEC), is expected to be more useful than the averaged null energy 
condition 
to put restrictions on the boundary stress-energy tensors\footnote{For static $3$-dimensional boundaries there are other constaints on the behavior of the boundary stress-energy 
tensors~\cite{Hickling:2015tza,Fischetti:2016vfq}.}.

%%%%%%%%%%%%%%%%%%%%%%%%%%%%%%%%%%%%%%%
\section{$d=5$ dimensional spatially compact universe}
\label{sec:5} 
%%%%%%%%%%%%%%%%%%%%%%%%%%%%%%%%%%%%%%%
In this section, we shall restrict our attention to $d=5$ static spatially compact boundary spacetime. 
In this case,  
the renormalized stress-energy tensor~(\ref{renormalized_stress_energy}) is determined by the $g_{(5)\mu \nu}$ term, which is at $O(\epsilon^5)$ and sub-dominant compared with 
the $g_{(4)\mu \nu}$ term at $O(\epsilon^4)$ in the expansion~(\ref{Fefferman-Graham_expansion}). 
Therefore the $g_{(4)\mu \nu}$ term determines whether or not the bulk-shortcut property is satisfied in the bulk.
We will show below that, up to $O(\epsilon^4)$, $U(\lambda_+)$ at the north pole for a bulk causal curve near the boundary null geodesic {\it cannot} be negative even when  
the boundary spacetime is largely deformed from the static Einstein compact universe. 
This implies that bulk-shortcut property is satisfied as long as the boundary CANEC, which is determined holographically by the $g_{(5)\mu \nu}$ term at $O(\epsilon^5)$, is satisfied.

%%%%%%%%%%%%%%%%%%%%%%%%%%%%%  
\subsection{General formulas}
%%%%%%%%%%%%%%%%%%%%%%%%%%%%%
Let us consider, as conformal boundary metric ${g}_{(0) \mu\nu}$, static spatially compact spacetime with the metric in double null coordinates 
\begin{align}
\label{double-null_higher}
ds^2_5=-2e^{\psi(\rho)}dUdV+2f^2(\rho)h_{mn}dx^m dx^n,  
\end{align}
where $h_{mn}dx^m dx^n$ are the metric of the three-dimensional unit sphere and the 
null coordinates are given by $U=t-\rho, \, V=t+\rho$.\footnote{In this section, we use slightly different notation compared with Sec. 3 and Sec. 4. There is a factor $\sqrt{2}$ difference for $U, V, t, r$ relationship and our conformal boundary metric \eqref{double-null_higher} is scaled by overall factor 2.} We assume that the south and north poles are located 
at $\rho=0$ and $\rho=\pi$~($0\le \rho \le \pi$), as before. Therefore when $t=\mbox{const}$. hypersurface becomes 
$4$-dimensional sphere, $f=\sin\rho$.   

Following the procedure in Sec. \ref{sec:3}, we consider a future-directed boundary null geodesic $\gamma$ along $U=0$ and $x^m=0$ 
from the south pole point $p$ to the north pole point $q$ with the tangent vector $l=e^{-\psi} {\partial_V}$.  
As before, the values of the affine parameter at $p$, $q$, are denoted by $\lambda_-$, $\lambda_+$, respectively, 
and the value of the null coordinate $V$ at $q$ is denoted by $V_0$. This null geodesic segment is achronal on the boundary, due to the spherical symmetry. 
%%%

Let us consider the bulk causal curves $\xi$ near the null geodesic $\gamma$ which connects $p$ and a point $r$ on the null line with 
tangent vector $\p_U$ through $q$. From the spherical symmetry, it is sufficient to examine $\xi$ with the tangent vector $K$
\begin{align}
K^a=\left(\frac{dz}{d\lambda},\, \frac{dU}{d\lambda},\,\frac{dV}{d\lambda},\,\frac{dx^m}{d\lambda} \right)
=(\dot{z}, K^U, e^{-\psi}, {\bm 0}).   
\end{align} 
$K^U$ and $z$ are expanded as a series in $\epsilon$ as 
\begin{align}
K^U=\epsilon^2 \frac{d{u_2}}{d\lambda}+\epsilon^3 \frac{d{u_3}}{d\lambda}+\epsilon^4 \frac{d{u_4}}{d\lambda}+\cdots, \quad z=\epsilon z_1+\epsilon^2 z_2+\epsilon^3 z_3+\cdots.   
\end{align}
Then, the condition $\hat{g}_{ab}K^a K^b\le 0$ can be expanded as 
\begin{align}
\label{Inequality:causal_full}
 0\ge \hat{g}_{ab}K^a K^b &=\epsilon^2\left(-2\frac{d{u_2}}{d\lambda}+\dot{z}_1^2
 +z_1^2g_{(2)\mu\nu}l^\mu l^\nu \right) \nonumber \\
&\quad +2\epsilon^3 \left(\dot{z}_1\dot{z}_2+z_1z_2\,g_{(2)\mu\nu}l^\mu l^\nu-\frac{d{u_3}}{d\lambda}  \right) \nonumber \\
&\,\, \quad +\epsilon^4\Biggl(\dot{z}_2^2+z_2^2\,g_{(2)\mu\nu}l^\mu l^\nu+
2\dot{z}_1\dot{z}_3+2z_1z_3\,g_{(2)\mu\nu}l^\mu l^\nu \nonumber \\
&\qquad \qquad+2z_1^2e^{-\psi}g_{(2)UV}\frac{d{u_2}}{d\lambda}-2\frac{d{u_4}}{d\lambda}
+z_1^4\,g_{(4)\mu\nu}l^\mu l^\nu   \nonumber \\
&\qquad \qquad \, + z_1^2 {  \partial_U \left(g_{(2)\mu\nu}  l^\mu l^\nu\right)} u_2  \Biggr) \,, 
\end{align}
%%%%
The last term is originated from the Taylor expansion for the metric $g_{(2)\mu\nu}$ near $U=0$, 
\begin{align}
& g_{(2)\mu\nu}l^\mu l^\nu(U, V) \nonumber \\
&= g_{(2)\mu\nu}l^\mu l^\nu (U=0, V) + U \partial_U \left(g_{(2)\mu\nu}l^\mu l^\nu\right) (U=0, V)  + O(U^2)  \,, 
\end{align}
and $u_2 = u_2 (\lambda)$ is given by 
\begin{align}
\label{Eq:u_2_def}
u_2(\lambda) = \frac{1}{2}\int^{\lambda}_{\lambda_-}\left(\dot{z}_1^2+z_1^2\,g_{(2)\mu\nu}l^\mu l^\nu \right)d\lambda \,. 
\end{align}
All metric functions should be evaluated at $U=0$ in the above formula. 
%%%%

Integrating the above inequality from $\lambda_-$ to $\lambda_+$, we obtain 
\begin{align}
& U(\lambda_+)=\int^{\lambda_+}_{\lambda_-} K^U d\lambda\ge {\cal I}, 
\end{align}
where ${\cal I}$ is expanded as ${\cal I}=\epsilon^2{\cal I}_2+\epsilon^3{\cal I}_3+\cdots$, and ${\cal I}_i$ are 
defined by  
\begin{align}
& {\cal I}_2:=\frac{1}{2}\int^{\lambda_+}_{\lambda_-}\left(\dot{z}_1^2+z_1^2\,g_{(2)\mu\nu}l^\mu l^\nu \right)d\lambda, \nonumber \\
& {\cal I}_3:= \int^{\lambda_+}_{\lambda_-}\left(\dot{z}_1\dot{z}_2+z_1z_2\,g_{(2)\mu\nu}l^\mu l^\nu \right)d\lambda, 
\nonumber \\
& {\cal I}_4:=\frac{1}{2}\int^{\lambda_+}_{\lambda_-} \biggl( \dot{z}_2^2+z_2^2\,g_{(2)\mu\nu}l^\mu l^\nu+
2\dot{z}_1\dot{z}_3+2z_1z_3\,g_{(2)\mu\nu}l^\mu l^\nu \nonumber \\
&\qquad \qquad \qquad +2z_1^2e^{-\psi}g_{(2)UV}\frac{d{u_2}}{d\lambda}+z_1^4\,g_{(4)\mu\nu}l^\mu l^\nu   
\nonumber \\
&\qquad \qquad \qquad \quad + { z_1^2  \partial_U \left(g_{(2)\mu\nu}  l^\mu l^\nu\right)} u_2 
\biggr)d\lambda. 
\end{align}

Note that $u_2(\lambda = \lambda_+) = {\cal I}_2$. 

Both ${\cal I}_2$ and ${\cal I}_3$ are minimized and become zero when $z_1$ satisfies 
\begin{align}
\label{Eq:z_1_higher}
\ddot{z}_1=g_{(2)\mu\nu}l^\mu l^\nu\,z_1=-\frac{1}{3}\hat{R}_{\mu\nu}l^\mu l^\nu\,z_1.  
\end{align}
Similarly, ${\cal I}_4$ is minimized when $z_2$ satisfies 
\begin{align}
\label{Eq:z_2_higher}
\ddot{z}_2=g_{(2)\mu\nu}l^\mu l^\nu\,z_2=-\frac{1}{3}\hat{R}_{\mu\nu}l^\mu l^\nu\,z_2, 
\end{align}
and reduces to  
\begin{align}
\label{Delta_U_general}
{\cal I}_4=\int^{\lambda_+}_{\lambda_-}\Bigl(z_1^2e^{-\psi}g_{(2)UV}\frac{d{u_2}}{d\lambda}
+\frac{z_1^4}{2}g_{(4)\mu\nu}l^\mu l^\nu  + { \frac{z_1^2}{2}  \partial_U\left( g_{(2)\mu\nu}  l^\mu l^\nu \right)}u_2
\Bigr)d\lambda, 
\end{align}  
where we used Eqs.~(\ref{Eq:z_1_higher}) and (\ref{Eq:z_2_higher}) and integration by part. 
This equation can be further simplified by noting that 
\begin{align}
u_2(\lambda) %= \frac{1}{2}\int^{\lambda}_{\lambda_-}\left(\dot{z}_1^2+z_1^2\,g_{(2)\mu\nu}l^\mu l^\nu \right)d\lambda \,. 
 =  \frac{1}{2}\int^{\lambda}_{\lambda_-}\left(\dot{z}_1^2 + z_1 \ddot{z}_1 \right)d\lambda 
 = \frac{1}{2} z_1 \dot{z}_1 
 \,, 
\end{align}
where we used Eqs.~(\ref{Eq:u_2_def}), (\ref{Eq:z_1_higher}) and integration by part with $z_1 (\lambda_-) = 0$. 
As a result, we obtain 
\begin{align}
\label{Delta_U_general}
U(\lambda_+)&=\epsilon^4\int^{\lambda_+}_{\lambda_-} \Bigl(  z_1^2e^{-\psi}g_{(2)UV}\frac{d{u_2}}{d\lambda}
+\frac{z_1^4}{2}g_{(4)\mu\nu}l^\mu l^\nu   \nonumber \\
&
\qquad  \qquad \qquad + \frac{\dot{z}_1 z^3_1}{4} \partial_U g_{(2)\mu\nu}l^\mu l^\nu 
\Bigr) d\lambda+O(\epsilon^5) \nonumber \\
& = \epsilon^4\int^{\lambda_+}_{\lambda_-} \Bigl(  z_1^2e^{-\psi}g_{(2)UV}\frac{d{u_2}}{d\lambda}
+\frac{z_1^4}{2}g_{(4)\mu\nu}l^\mu l^\nu   \nonumber \\
& \qquad  \qquad \qquad + \frac{ z^4_1}{16} { \frac{d \{\partial_V \left(g_{(2)\mu\nu}l^\mu l^\nu \right)\}}{d \lambda}} 
\Bigr) d\lambda+O(\epsilon^5) 
\end{align}  
where in the last equality, we used the static nature of the boundary metric, 
\begin{align}
\partial_U (g_{(2)\mu\nu}l^\mu l^\nu)  = -  \partial_V (g_{(2)\mu\nu}l^\mu l^\nu) 
\end{align}
and integration by part. 
%%%

Note that in the static Einstein universe with the metric $\psi=0$, $f(\rho)=\sin \rho$, 
$g_{(2)UV} = g_{(4)\mu\nu}l^\mu l^\nu  =  \partial_U \left(g_{(2)\mu\nu} l^\mu l^\nu \right) =0$, therefore $U(\lambda_+)=0$, up to $O(\epsilon^4)$. 
This means that $r$ is equal to $q$, up to $O(\epsilon^4)$. In this case, focusing on $O(\epsilon^5)$, we find 
\begin{align}
\label{order5thbulkcondition}
 0\ge \hat{g}_{ab}K^a K^b &=\epsilon^5 \biggl(  2 \dot{z_2} \dot{z_3} + 2 \dot{z_1} \dot{z_4}  +   2 {z_2} {z_3} g_{(2)\mu\nu}\,l^\mu l^\nu  + 2 {z_1} {z_4} g_{(2)\mu\nu}\,l^\mu l^\nu   \nonumber \\
 & \qquad \qquad   +   4 z_1^3 z_2 g_{(4)\mu\nu}\,l^\mu l^\nu
+ z_1^5\,g_{(5)\mu\nu}\,l^\mu l^\nu  \nonumber \\
& \qquad \qquad \,\,  - 2 \frac{d u_5}{d \lambda} + 4 z_1 z_2e^{-\psi}g_{(2)UV}\frac{d{u_2}}{d\lambda} \nonumber \\
& \qquad \qquad \quad + 2 z_1 z_2  {  \partial_U\left(g_{(2)\mu\nu}l^\mu l^\nu\right)}  u_2 \biggr) \,.
\end{align}
The last term is originated from the Taylor expansion for the metric $g_{(2)\mu\nu}$ near $U=0$.  
By using \eqref{Eq:z_1_higher} and \eqref{Eq:z_2_higher} and $g_{(2)UV}  =g_{(4)\mu\nu}l^\mu l^\nu =  \partial_U \left(g_{(2)\mu\nu} l^\mu l^\nu \right) =0$, 
we obtain the following null inequality
\begin{align}
\label{Null_inequality_5}
\int^{\lambda_+}_{\lambda_-}z_1^5\,g_{(5)\mu\nu}\,l^\mu l^\nu d\lambda\ge 0,  
\end{align}
as obtained in Sec. \ref{sec:3}. The 5-dimensional Raychaudhuri equation for the null geodesic congruences of the boundary null 
geodesic $\gamma$ is written by the expansion $\theta$ as  
\begin{align}
\frac{d\theta}{d\lambda}=-\frac{\theta^2}{3}  -\hat{R}_{\mu\nu}l^\mu l^\nu=-\frac{\theta^2}{3}-\frac{3}{4}. 
\end{align} 
By defining $\theta=d(\eta^3)/d\lambda/\eta^3$, it becomes the Jacobi equation 
\begin{align}
\frac{d^2\eta}{d\lambda^2}=-\frac{1}{3}\hat{R}_{\mu\nu}l^\mu l^\nu\,\eta=-\frac{\eta}{4}. 
\end{align}
This is equivalent to Eq.~(\ref{Eq:z_1_higher}), implying that $z_1$ can be replaced by the Jacobi field $\eta$ in the boundary 
theory, as shown in the $d=3$ case, and we obtain holographic boundary conformally invariant averaged null energy condition (CANEC)  
\begin{align}
\label{Null_inequality_5}
\int^{\lambda_+}_{\lambda_-} \eta^5\, \braket{T_{\mu\nu}} \,l^\mu l^\nu d\lambda\ge 0 .
\end{align}
The conformal invariance of the above formula can easily been seen by using the formulae, 
$ \tilde{l}^\mu={\Omega^{-2}}l^\mu ,   \, {d\tilde{\lambda}} =\Omega^2  {d\lambda}, \, \tilde{\eta}=\Omega  \, \eta$
as given in Eq.~(\ref{conformal_t}), and noting that in $5$-dimensions, 
\begin{align}
\langle {\tilde T}_{\mu \nu} \rangle = \Omega^{-3} \langle {T}_{\mu \nu} \rangle . 
\end{align}

%%%%%%%%%%%%%%%%%%%%%%%%%%%%%%%%%%%
\subsection{$d=5$ CANEC}
%%%%%%%%%%%%%%%%%%%%%%%%%%%%%%%%%%%
%%%%%o%%%%%%%%%%%

\begin{figure}[tbp]
\begin{center}
\includegraphics[width=.6\textwidth]{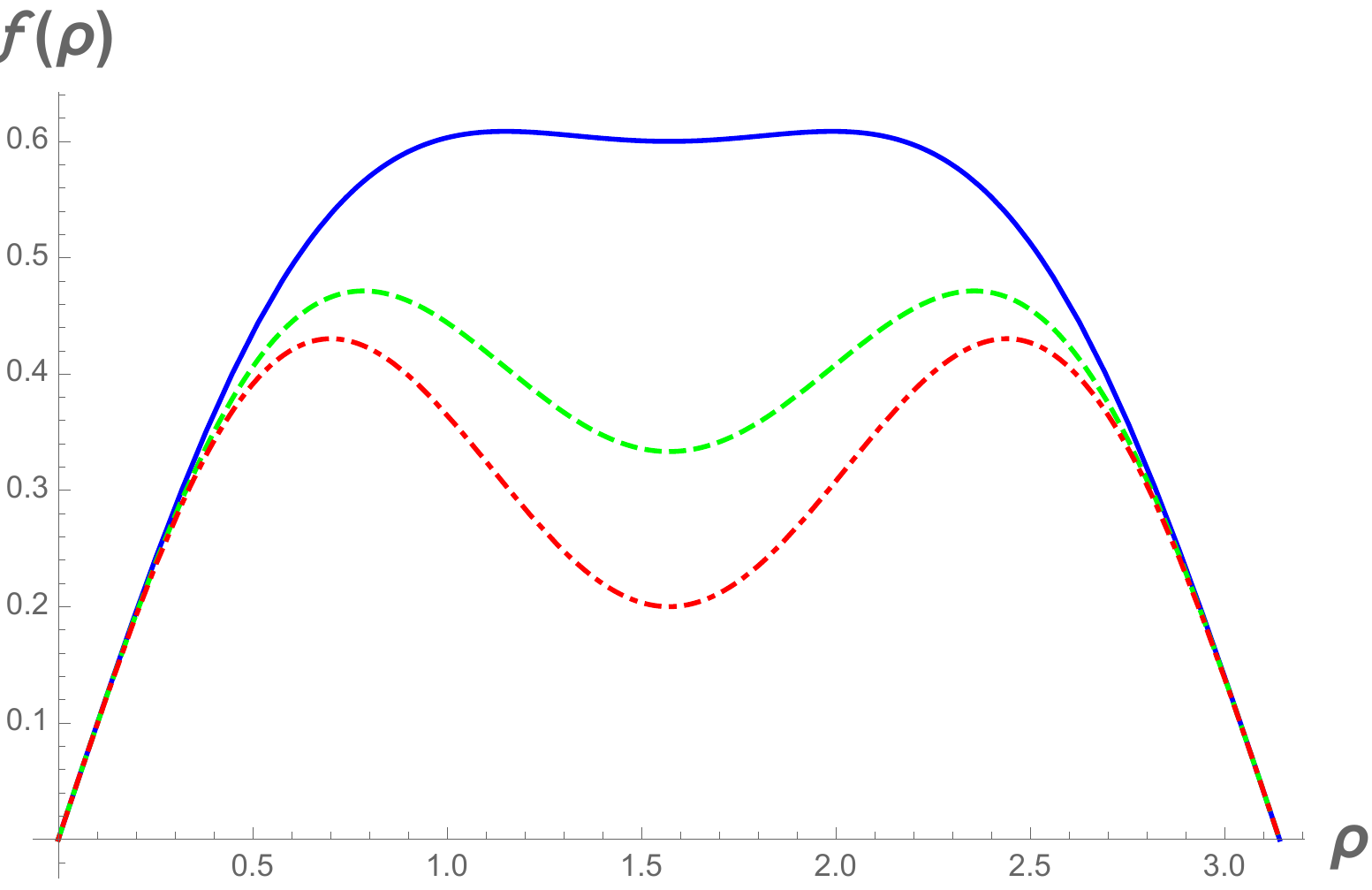}
\vspace{5mm}
\caption{\label{g_function} The function $f(\rho)$ is plotted for $\delta=1/5$~(solid, blue), $1/3$~(dashed, green), $2/5$~(dot-dashed, red).}
\end{center}
\end{figure}

Let us now turn to the case of deformed static Einstein universe as our $5$-dimensional boundary. 
We consider the function $\psi$ and $f$ of the metric \eqref{double-null_higher} as  
\begin{align}
\psi(\rho) &= 0 \,, \nonumber \\
f(\rho)&=(1-\delta+\delta \cos 2\rho)\sin\rho\,, \qquad 0\le \delta<\frac{1}{2} \,.
\end{align}
The function is plotted in Fig.~\ref{g_function} for various values of $\delta$. When $\delta=0$, the function reduces to 
$f=\sin\rho$, and the metric~(\ref{double-null_higher}) becomes the static Einstein universe. As $\delta$ increases, the compact spatial 
sections are more and more deformed to approach ``dumbbell" like shape. 

%%%%%%%%%%%%

In this case, the first term in Eq.~(\ref{Delta_U_general}) becomes 
\begin{align}
\label{du_2_higher}
& \int^{2\pi}_{0}z_1^2e^{-\psi}g_{(2)UV}\frac{d{u_2}}{d\lambda}d\lambda \nonumber \\
&=\frac{1}{16}\int^{2\pi}_{0}z_1^2\frac{f'^2-1-ff''}{f^2}\left(\dot{z_1}^2+\frac{f''}{4f}z_1^2 \right)d\lambda \nonumber \\
&=\frac{1}{768}\int^{2\pi}_{0}\frac{z_1^4}{f^4}\left[6f'^4-2f'^2(3+2ff'')+4f^2f'f^{(3)}-f\{f''(6+5ff'')+f^2f^{(4)}\}\right]d\lambda, 
\end{align}
where $f^{(n)}$ is the $n$-th derivative of the function with respect to $\rho$, and we set $\lambda_-=0$. 
To derive the above equation, we used $d\lambda=dV=2d\rho$, therefore $\rho = \pi$ corresponds to $\lambda_+ = 2 \pi$.  We also used  
\begin{align}
\frac{d{u_2}}{d\lambda}=\frac{1}{2}\left(\dot{z}_1^2+z_1^2g_{(2)\mu\nu}l^\mu l^\nu \right)
\end{align}
for the bulk null curve $\xi$ in the second equality. To derive the last equality, we used 
integration by part under the condition~(\ref{Eq:z_1_higher}). 
Substituting Eq.~(\ref{du_2_higher}) into Eq.~(\ref{Delta_U_general}) and noting 
\be
{ \frac{1}{16}\frac{d \left\{ \partial_V (g_{(2)\mu\nu} l^\mu l^\nu)\right\}}{d \lambda}} 
= \frac{1}{16}\frac{d \left( \partial_V g_{(2)\mu\nu} \right)}{d \lambda} l^\mu l^\nu 
 = \frac{1}{64} \frac{d^2 g_{(2)VV} }{d \rho^2} = \frac{1}{64} \frac{d^2 }{d \rho^2} \left( \frac{f''}{4f} \right)
\ee 
and  
\begin{align}
\label{coefficient_4_VV}
 g_{(4)\mu\nu}\,l^\mu l^\nu  
= \frac{1}{128 f^4} 
\left[
   -2f'^4 + f'^2 (2-4 f f'')+ f f''(2+ff'')+4f^2f' f^{(3)}+ f^3 f^{(4)}
\right] , 
\end{align}
for the constant vector $l^\mu$ in this case, 
we obtain 
\begin{align}
\label{Delta_U_pre_final_higher}
U(\lambda=2\pi)&=\frac{5 \epsilon^4}{768}\int^{2\pi}_{0}\frac{z_1^4}{f^3}
\left[-2 f'^2f''+2 ff'f^{(3)}+f(-f''^2+ff^{(4)})   \right]d\lambda  +O(\epsilon^5) \nonumber \\ 
& = \frac{5 \epsilon^4}{192}\int^{2\pi}_{0} {z_1^4} \left[ \frac{d^2 g_{(2)VV} }{d \rho^2} + \frac{4 f'}{f} \frac{d g_{(2)VV} }{d \rho}\right]d\lambda +O(\epsilon^5)
\end{align}   
for the bulk null curve $\xi$. 
Furthermore, Eq.~(\ref{Eq:z_1_higher}) becomes 
\begin{align}
\frac{d^2 z_1}{d \rho^2}   = \frac{f''}{f} z_1. 
\end{align} 
Since $f$ satisfies the same boundary condition as $z_1$, i.~e.~$f(0)=f(\pi)=0$, and (\ref{Eq:z_1_higher}) is linear with 
respect to $z_1$, $z_1$ is proportional to $f$.  We set its proportional coefficient as $\alpha_1$, 
\begin{align}
\label{alphaone}
z_1 = \alpha_1 f .
\end{align}
Plugging this into Eq.~(\ref{Delta_U_pre_final_higher}) and using the boundary conditions $f(0) = f(\pi) = 0$, we find that $U(\lambda=2 \pi)$ vanishes, up to $O(\epsilon^4)$, 
\begin{align}
\label{Delta_U_final_higher}
U(\lambda=2\pi)
& = \frac{5 \epsilon^4 \alpha_1^4}{96}\int^{\pi}_{0}  \frac{d}{d \rho} \left[f^4  \frac{d g_{(2)VV} }{d \rho} \right]   d\rho +O(\epsilon^5) = O(\epsilon^5) \,. 
\end{align}
This implies that whether the no-bulk-shortcut property is satisfied or not is determined by $O(\epsilon^5)$. 
It is straightforward to show that in fact, at $O(\epsilon^5)$ 
the no-bulk-shortcut property leads exactly to the CANEC as we have seen in the previous subsection. 
To show this,  in addition to Eq.~(\ref{Eq:z_1_higher}) and (\ref{Eq:z_2_higher}),  
by setting $z_2$ also as 
\begin{align}
z_2 = \alpha_2 f, 
\end{align} 
we use the following relation    
\begin{align}
&\int_0^{2 \pi} d\lambda \left(   4 z_1^3 z_2 g_{(4)\mu\nu}\,l^\mu l^\nu  
 + 4 z_1 z_2e^{-\psi}g_{(2)UV}\frac{d{u_2}}{d\lambda}  \right) \nonumber \\
 & \quad =  \frac{\alpha_1^3 \alpha_2}{4} \int_0^{2 \pi} d\lambda \left( f^3 \frac{df}{d\rho}   \left( \partial_\rho g_{(2) VV} \right) \right) ,
\end{align}  
under substitution of (\ref{coefficient_4_VV}) 
and also use 
\begin{align}
& \int_0^{2 \pi} d\lambda \left(  2 z_1 z_2  \left( \partial_U g_{(2)\mu\nu} \right) l^\mu l^\nu u_2 \right)   
= - \int_0^{2 \pi} d\lambda z_1^2 \dot{z}_1 z_2  \left( \partial_V g_{(2) VV} \right)    \nonumber \\
& %= - \alpha_1^3 \alpha_2 \int_0^{2 \pi} d\lambda  \left( f^3 \dot{f}   \left( \partial_V g_{(2) VV} \right)  \right)
\qquad  \qquad = - \frac{\alpha_1^3 \alpha_2}{4} \int_0^{2 \pi} d\lambda \left( f^3 \frac{df}{d\rho}   \left( \partial_\rho g_{(2) VV} \right) \right) . 
\end{align}

In summary, we have seen that the resultant bulk satisfies the no-bulk-shortcut property 
as far as CANEC for the dual boundary theory is satisfied.

%%%%%%%%%%%%%%%%
\section{Summary}
\label{sec:summary}
%%%%%%%%%%%%%%%%
In this paper, we have pursued the question of what is a sensible causal interaction between the bulk and conformal boundary in the 
context of the AdS/CFT duality, by regarding the Gao-Wald's no-bulk-shortcut property as our guiding principle for the appropriate choice of the boundary metric. 
In Sec.~\ref{sec:3} we have considered $3$-dimensional conformal boundary whose spatial section is either positively or negatively curved. For the positive curvature case, by examining holographic bulk causal curves close to a certain boundary null geodesic, we have derived the conformally invariant averaged null energy condition (CANEC) which can be used to put a certain restriction on the extent of the negative value of 
the null energy for a strongly coupled field theory. For the negatively curved case, 
the no-bulk-shortcut property is automatically satisfied for the bulk causal curves near the boundary null geodesic, and hence, we cannot derive the null energy inequality. 
In Sec.~\ref{sec:4}, we have generalized the CANEC obtained in Sec.~\ref{sec:3} to the case of a deformed compact boundary. Then we have shown the conformal invariance of the CANEC.  
In Sec.~\ref{sec:5}, we have considered a $5$-dimensional deformed static Einstein universe as our boundary geometry, and 
obtained CANEC for the $5$-dimensional boundary. 

In all of the examples we have studied in this paper, the no-bulk-shortcut property implies that CANEC holds on the boundary. 
This, in turn, implies that if the boundary CANEC is violated,  
then there must exist a bulk-shortcut, hence giving rise to a mismatch between the bulk and boundary correlators. 
Therefore we expect that {\it a boundary field theory violating CANEC cannot have a consistent holographic bulk dual.}   

Our results are based on the Fefferman-Graham expansion from the given boundary geometry and also on the holographic stress-energy formulas of 
Ref.~\cite{deHaro:2000vlm}. As one of the future directions, it would be quite interesting to examine whether there are some concrete examples violating 
no-bulk-shortcut property in $d=5$. In Sec.~\ref{sec:5}, we have assumed that the boundary spacetime is restricted to a class of the {\it static} spacetime. 
If this assumption is relaxed, it is non-trivial whether the no-bulk-shortcut property is satisfied at $O(\epsilon^4)$, due to the existence of the 
higher curvature terms~(\ref{coeff_4}).    
Another issue left open for future work is the even-dimensional boundary case, which involves conformal anomalies and as a result, things are more complicated.  
It would be interesting to see whether, and how, the presence of conformal anomalies could affect the causality aspects of bulk and boundary duality.   

Finally we discuss what type of boundary metrics can admit the Einstein gravity as the holographic bulk dual. 
From the bulk viewpoint, given boundary metric as an ``initial condition'', one can solve the bulk Einstein equations as radial $z$ evolution. From this perspective, one might think that any boundary metric can be %OK.
acceptable in the AdS/CFT context. However, although all of our boundary metrics examined in this paper lead to CANEC under the no-shortcut property, it is still unclear if there is an example of boundary metric which does not allow the bulk no-shortcut property to hold whatever the boundary stress tensor we have chosen.  
For example, if there is a boundary metric which leads to the violation of the bulk no-shortcut property at $O(\epsilon^4)$ for $d=5$ case due to the existence of the higher curvature terms~(\ref{coeff_4}), %as mentioned before, then  
the dual bulk must then form a naked singularity visible from the boundary, hence violating the weak cosmic censorship, according to the result of~\cite{Ishibashi:2019nby}. This is independent of 
wherther CANEC holds or not, because CANEC for $d=5$ case is determined  at subleading order, $O(\epsilon^5)$.  
It would be interesting to investigate the weak cosmic censorship in more detail, from the causal consistency viewpoint of the AdS/CFT duality.

\bigskip
\goodbreak
\centerline{\bf Acknowledgments}
\noindent
We would like to thank Gary Horowitz for useful comments on the draft.
This work was supported in part by JSPS KAKENHI Grant No. 18K03619 (N.I.), 15K05092 (A.I.), 17K05451 (K.M.).

\appendix
%%%%%%%%%%%%%%%%%%%%%%%%%%%%%%%%%%%%%%%%%%%%%%%%%%%%%%%%%%%%%%%%%%%%%%%%%%%%%%
\section{$d=4$ static Einstein spatially compact universe}
%%%%%%%%%%%%%%%%%%%%%%%%%%%%%%%%%%%%%%%%%%%%%%%%%%%%%%%%%%%%%%%%%%%%%%%%%%%%%%

In this appendix, we consider the following $4$-dimensional boundary spacetime whose spatial section is a constant curvature space, 
\begin{align} 
\label{def:4bm}
& ds_4^2=-dt^2+d\rho^2  + f^2_{k}(\rho) (d\theta^2+\sin^2\theta d\varphi^2) \,,  
\end{align}
where 
$$ 
f_k (\rho) = \frac{1}{\sqrt{k}} \sin \sqrt{k} \rho \,, \quad k= \pm 1, \,0 \,, 
$$   
so that $k$ describes the normalized sectional curvature of the $t=const. $ hypersurfaces. 
In particular, when $k=0$, the boundary spacetime with the above metric becomes $4$-dimensional Minkowski spacetime, 
and when $k=+1$, it describes $4$-dimensional Einstein-static universe. 
We introduce the double null coordinates as
$$
    U=\frac{t-\rho}{\sqrt{2}}, \qquad V=\frac{t+\rho}{\sqrt{2}} . 
$$
The stress-energy tensor on the boundary spacetime is given by 
\begin{align}
 \braket{T_{\mu\nu}}=g_{(4)\mu\nu}-\frac{1}{8}g_{(0)\mu\nu}\left[(g^\alpha_{(2)\alpha})^2-g^\alpha_{(2)\beta}g^\beta_{(2)\alpha} \right]
-\frac{1}{2}g_{(2)\mu\alpha}g^\alpha_{(2)\nu}+\frac{1}{4}g_{(2)\mu\nu}g^\alpha_{(2)\alpha}, 
\end{align}  
where we set $4\pi G_4=1$ and the indices are raised and lowered with the leading metric $g_{(0)\mu\nu}$~\cite{deHaro:2000vlm}, which is given by Eq.~(\ref{def:4bm}) in the present case. Then, we obtain the null-null component of the stress-energy tensor as 
\begin{align}
\label{stress-energy_4}
 \braket{T_{\mu\nu}}l^\mu l^\nu=g_{(4)\mu\nu}l^\mu l^\nu+\frac{k^2}{8}, \qquad l=\p_V. 
\end{align} 
We can immediately see the difference between the Minkowski boundary case $k=0$, which is considered in \cite{Kelly:2014mra}, 
and the curved boundary case $k =\pm 1$, which involves the gravitational anomaly term in the right-hand side. 
We also note that the effects of the curvature on the stress-energy tensor are the same in both the positive 
and negative curvature case. However, when we consider non-local effects by integrating the stress-energy 
component (\ref{stress-energy_4}) along the null geodesic with the tangent $l^\mu$ as considered in the previous sections, 
we have to take into account the global structure of the curved boundary. 
For the positive curvature $k=+1$ case, the boundary is a static closed universe 
and therefore the null geodesic naturally admits a pair of conjugate points on it (at the north and south poles), which makes the null geodesic fail to be achronal. In contrast, for the negative curvature $k=-1$ case, the boundary is a static open-universe and the null geodesic with $l^\mu$ does not admit a pair of conjugate points, unless some non-trivial compactification is made. 
 
In what follows, we consider the positive curvature $k=+1$ case. 
As done in Sec. \ref{sec:3}, expanding the tangent vector $K$ as 
\begin{align}
\label{bulk_causal_curve_4}
 K^a &=\left(\frac{dz}{d\lambda},\, \frac{dU}{d\lambda},\,\frac{dV}{d\lambda},\,0,\, 0 \right)=
\left(\frac{dz(\lambda)}{d\lambda}, \, K^U,\, 1,\, 0,\, 0\right), \nonumber \\
 z &=\epsilon z_1+\epsilon^2 z_2 +\cdots, \nonumber \\
 K^U &=\epsilon^2 \frac{du_2}{d\lambda}+\epsilon^3 \frac{du_3}{d\lambda}+\cdots.  
\end{align}
and imposing $\hat{g}_{ab}K^a K^b\le 0$, one obtains the same inequality~Eq.~(\ref{u2_IE_positive_con})
at $O(\epsilon^2)$. So, the curve satisfying Eq.~(\ref{sol:z_1_positive_con}) gives the minimum of ${\cal I}$, 
and $U$ at $r$ is zero at $O(\epsilon^2)$.  

Similarly, one derives the following inequality at $O(\epsilon^4)$ as  
\begin{align}
\frac{du_4}{d\lambda}\ge \frac{1}{2}\left(\dot{z_2}^2-\frac{1}{2}z_2^2+2\dot{z_1}\dot{z_3}-z_1z_3
+z_1^4\, \left( g_{(4)\mu\nu}  + \ln z^2 h_{\mu\nu} \right) \,l^\mu l^\nu\right) \,,
\end{align}
One can check that for the constant curvature space with the metric given by Eq.~\eqref{def:4bm}, 
\be
h_{\mu\nu} = 0 \,,
\ee 
where $h_{\mu\nu}$ in Eq.~\eqref{Fefferman-Graham_expansion} is given by Eq.~(A.6) of \cite{deHaro:2000vlm}\footnote{Again take care of the Riemann tensor notation difference.}.  
By using the integration by part under the boundary condition $z_3(0)=z_3(\sqrt{2}\pi)=0$, one finds that 
\begin{align}
\label{U4_positive_con}
U(\sqrt{2}\pi)\ge \frac{\epsilon^4}{2} \int^{\sqrt{2}\pi}_{0}z_1^4\,g_{(4) \mu\nu}\,l^\mu l^\nu d\lambda  
=\frac{\epsilon^4}{2} \int^{\sqrt{2}\pi}_{0}\left(\sin\frac{\lambda}{\sqrt{2}}\right)^4\,g_{{(4)}\mu\nu}\,l^\mu l^\nu d\lambda
\ge 0 \,. 
\end{align}
Combining this with Eq.~(\ref{stress-energy_4}), we find the inequality for the null-null component of the 
stress-energy tensor as 
\begin{align}
\int^{\sqrt{2}\pi}_{0}z_1^4\, \braket{T_{\mu\nu}}l^\mu l^\nu\, d\lambda\ge \frac{1}{8}\int^{\sqrt{2}\pi}_{0}z_1^4 d\lambda>0 \,. 
\end{align}

%%%%%%%%%%%%%%%%%%%%%%%%%%%%%%%%%%%%%%%%%%%%%%%%%%%%%%%%%%%%%%%%%%%%%%%%%%%%%%%%


\begin{thebibliography}{99}
%%%%%%%%%%%%%%%%%%%%%%%%%%%%%%%%%%%%%%%%%%%%%%%%%%%%%%%%%%%%%%%%%%%%%%%%%%%%%%%%
%\cite{Maldacena:1997re}
\bibitem{Maldacena:1997re}
  J.~M.~Maldacena,
  ``The Large N limit of superconformal field theories and supergravity,''
  Int.\ J.\ Theor.\ Phys.\  {\bf 38}, 1113 (1999)
  [Adv.\ Theor.\ Math.\ Phys.\  {\bf 2}, 231 (1998)]
  doi:10.1023/A:1026654312961, 10.4310/ATMP.1998.v2.n2.a1
  [hep-th/9711200].

%\cite{Gubser:1998bc}
\bibitem{Gubser:1998bc} 
  S.~S.~Gubser, I.~R.~Klebanov and A.~M.~Polyakov,
  ``Gauge theory correlators from noncritical string theory,''
  Phys.\ Lett.\ B {\bf 428}, 105 (1998)
  doi:10.1016/S0370-2693(98)00377-3
  [hep-th/9802109].
  %%CITATION = doi:10.1016/S0370-2693(98)00377-3;%%
  %8176 citations counted in INSPIRE as of 08 Aug 2019

%\cite{Witten:1998qj}
\bibitem{Witten:1998qj} 
  E.~Witten,
  ``Anti-de Sitter space and holography,''
  Adv.\ Theor.\ Math.\ Phys.\  {\bf 2}, 253 (1998)
  doi:10.4310/ATMP.1998.v2.n2.a2
  [hep-th/9802150].
  %%CITATION = doi:10.4310/ATMP.1998.v2.n2.a2;%%
  %9578 citations counted in INSPIRE as of 08 Aug 2019

%\cite{Heemskerk:2009pn}
\bibitem{Heemskerk:2009pn} 
  I.~Heemskerk, J.~Penedones, J.~Polchinski and J.~Sully,
  ``Holography from Conformal Field Theory,''
  JHEP {\bf 0910}, 079 (2009)
  doi:10.1088/1126-6708/2009/10/079
  [arXiv:0907.0151 [hep-th]].
  %%CITATION = doi:10.1088/1126-6708/2009/10/079;%%
  %402 citations counted in INSPIRE as of 11 Aug 2019



%\cite{ElShowk:2011ag}
\bibitem{ElShowk:2011ag} 
  S.~El-Showk and K.~Papadodimas,
  ``Emergent Spacetime and Holographic CFTs,''
  JHEP {\bf 1210}, 106 (2012)
  doi:10.1007/JHEP10(2012)106
  [arXiv:1101.4163 [hep-th]].
  %%CITATION = doi:10.1007/JHEP10(2012)106;%%
  %151 citations counted in INSPIRE as of 11 Aug 2019


%\cite{Camanho:2014apa}
\bibitem{Camanho:2014apa} 
  X.~O.~Camanho, J.~D.~Edelstein, J.~Maldacena and A.~Zhiboedov,
  ``Causality Constraints on Corrections to the Graviton Three-Point Coupling,''
  JHEP {\bf 1602}, 020 (2016)
  doi:10.1007/JHEP02(2016)020
  [arXiv:1407.5597 [hep-th]].
  %%CITATION = doi:10.1007/JHEP02(2016)020;%%
  %298 citations counted in INSPIRE as of 28 Oct 2019


%\cite{Afkhami-Jeddi:2016ntf}
\bibitem{Afkhami-Jeddi:2016ntf} 
  N.~Afkhami-Jeddi, T.~Hartman, S.~Kundu and A.~Tajdini,
  ``Einstein gravity 3-point functions from conformal field theory,''
  JHEP {\bf 1712}, 049 (2017)
  doi:10.1007/JHEP12(2017)049
  [arXiv:1610.09378 [hep-th]].
  %%CITATION = doi:10.1007/JHEP12(2017)049;%%
  %51 citations counted in INSPIRE as of 28 Oct 2019

  
%\cite{Gao:2000ga}
\bibitem{Gao:2000ga} 
  S.~Gao and R.~M.~Wald,
  ``Theorems on gravitational time delay and related issues,''
  Class.\ Quant.\ Grav.\  {\bf 17}, 4999 (2000)
  doi:10.1088/0264-9381/17/24/305
  [gr-qc/0007021].
  %%CITATION = doi:10.1088/0264-9381/17/24/305;%%
  %123 citations counted in INSPIRE as of 18 Oct 2019
  
  

  
\bibitem{Witten:2019qhl} 
  E.~Witten,
  ``Light Rays, Singularities, and All That,''
  arXiv:1901.03928 [hep-th].
  %%CITATION = ARXIV:1901.03928;%%
  %6 citations counted in INSPIRE as of 02 Sep 2019     
  

\bibitem{Kelly:2014mra} 
  W.~R.~Kelly and A.~C.~Wall,
  ``Holographic proof of the averaged null energy condition,''
  Phys.\ Rev.\ D {\bf 90}, no. 10, 106003 (2014)
  Erratum: [Phys.\ Rev.\ D {\bf 91}, no. 6, 069902 (2015)]
  doi:10.1103/PhysRevD.90.106003, 10.1103/PhysRevD.91.069902
  [arXiv:1408.3566 [gr-qc]].




%\cite{Faulkner:2016mzt}
\bibitem{Faulkner:2016mzt} 
  T.~Faulkner, R.~G.~Leigh, O.~Parrikar and H.~Wang,
  ``Modular Hamiltonians for Deformed Half-Spaces and the Averaged Null Energy Condition,''
  JHEP {\bf 1609}, 038 (2016)
  doi:10.1007/JHEP09(2016)038
  [arXiv:1605.08072 [hep-th]].
  %%CITATION = doi:10.1007/JHEP09(2016)038;%%
  %105 citations counted in INSPIRE as of 11 Aug 2019


%\cite{Hartman:2016lgu}
\bibitem{Hartman:2016lgu} 
  T.~Hartman, S.~Kundu and A.~Tajdini,
  ``Averaged Null Energy Condition from Causality,''
  JHEP {\bf 1707}, 066 (2017)
  doi:10.1007/JHEP07(2017)066
  [arXiv:1610.05308 [hep-th]].
  %%CITATION = doi:10.1007/JHEP07(2017)066;%%
  %72 citations counted in INSPIRE as of 11 Aug 2019
  
  

\bibitem{Urban:2009yt} 
  D.~Urban and K.~D.~Olum,
  ``Averaged null energy condition violation in a conformally flat spacetime,''
  Phys.\ Rev.\ D {\bf 81}, 024039 (2010)
  doi:10.1103/PhysRevD.81.024039
  [arXiv:0910.5925 [gr-qc]].
 
\bibitem{Visser:1994jb}
  M.~Visser,
  ``Scale anomalies imply violation of the averaged null energy condition,''
  Phys.\ Lett.\ B {\bf 349}, 443 (1995)
  doi:10.1016/0370-2693(95)00303-3
  [gr-qc/9409043].


%\cite{Ishibashi:2019nby}
\bibitem{Ishibashi:2019nby} 
  A.~Ishibashi, K.~Maeda and E.~Mefford,
  ``Achronal averaged null energy condition, weak cosmic censorship, and AdS/CFT duality,''
  Phys.\ Rev.\ D {\bf 100}, no. 6, 066008 (2019)
  doi:10.1103/PhysRevD.100.066008
  [arXiv:1903.11806 [hep-th]].
  %%CITATION = doi:10.1103/PhysRevD.100.066008;%%
   
  
\bibitem{HawkingEllis}
S.~W.~Hawking and G.~F.~R.~Ellis, ``{\it The large scale structure of spacetime}", Cambridge University Press~(1973). 


\bibitem{deHaro:2000vlm} 
  S.~de Haro, S.~N.~Solodukhin and K.~Skenderis,
  ``Holographic reconstruction of space-time and renormalization in the AdS / CFT correspondence,''
  Commun.\ Math.\ Phys.\  {\bf 217}, 595 (2001)
  doi:10.1007/s002200100381
  [hep-th/0002230].
  
  
\bibitem{Wald}
R.~M.~Wald, ``{\it General Relativity}", The university of Chicago Press~(1984). 


\bibitem{Emparan:1999gf} 
  R.~Emparan,
  ``AdS / CFT duals of topological black holes and the entropy of zero energy states,''
  JHEP {\bf 9906}, 036 (1999)
  doi:10.1088/1126-6708/1999/06/036
  [hep-th/9906040].
  
  
\bibitem{Graham:2007va} 
  N.~Graham and K.~D.~Olum,
  ``Achronal averaged null energy condition,''
  Phys.\ Rev.\ D {\bf 76}, 064001 (2007)
  doi:10.1103/PhysRevD.76.064001
  [arXiv:0705.3193 [gr-qc]].  

 
\bibitem{Hickling:2015tza}
   A.~Hickling and T.~Wiseman,
  ``Vacuum energy is non-positive for (2 + 1)-dimensional holographic CFTs"
   Class. Quant. Grav. {\bf 33}, 045009 (2016)
   doi:10.1088/0264-9381/33/4/045009     
   [arXiv:1508.04460 [hep-th]].
 
 
\bibitem{Fischetti:2016vfq}
   S.~Fischetti, A.~Hickling and T.~Wiseman,
 ``Bounds on the local energy density of holographic CFTs from bulk geometry"
   Class. Quant. Grav. {\bf 33}, 225003 (2016)
   doi:10.1088/0264-9381/33/22/225003
   [arXiv:1605.00007 [hep-th]].
 



\end{thebibliography}
\end{document}